\documentclass[11pt]{article} 
\begin{document} 
  \vspace*{1.1cm} 
  \begin{center} 
  {\LARGE \bf Zero-point field induced mass \\ vs.\ QED
mass renormalization} 
  \end{center} 

  \begin{center} 
  \vskip 10pt 
  Giovanni Modanese\footnote{e-mail address: 
  giovanni.modanese@unibz.it}
  \vskip 5pt
  {\it California Institute for Physics and Astrophysics \\
  366 Cambridge Ave., Palo Alto, CA 94306}
  \vskip 5pt
  and
  \vskip 5pt
  {\it University of Bolzano -- Industrial Engineering \\
  Via Sernesi 1, 39100 Bolzano, Italy}

  \vskip 10pt

 To appear in the proceedings of the
18th Advanced ICFA Beam Dynamics Workshop on "Quantum Aspects of 
Beam Physics", Capri, Italy, October 15-20, 2000
  \end{center} 

  \baselineskip=.175in 

\begin{abstract}
Haisch and Rueda have recently proposed a model in which
the inertia of charged particles is a consequence of
their interaction with the electromagnetic zero-point field. This
model is based on the observation that in an accelerated
frame the momentum distribution of vacuum fluctuations is
not isotropic. We analyze this issue through standard
techniques of relativistic field theory, first by regarding
the field $A_\mu$ as a classical random field, and then
by making reference to the mass renormalization procedure
in Quantum Electrodynamics and scalar-QED.
\end{abstract}

A general property of vacuum fluctuations in any 
 relativistic theory is the Lorentz invariance of their 
 frequency spectrum. This is true for quantum field theories 
 and also for ``stochastic" field theories -- those models in 
 which the use of commuting fields supplemented by a 
 suitable stochasticity principle allows to re-obtain most 
 results of the full quantum field theories \cite{sto}. In any case, 
 it is required that all observers, independently of their 
 relative motion, see the same fluctuations spectrum. 
  
 However, it has been known for a long time (since the 
 work by Unruh and Davies \cite{ud}) 
 that an accelerated observer 
 will see a different spectrum of the vacuum fluctuations, 
 corresponding to an apparent non-zero temperature. The 
 temperature is a function of the acceleration: 
 $T_a = a/(2\pi ck)$. This effect 
 offers a method of principle for detecting absolute 
 acceleration. 
  
 Since the acceleration has a definite direction and versus, 
 one also expects that the accelerated observer will notice an 
 anisotropy in the distribution of the vacuum fluctuations. 
 For the electromagnetic field such anisotropy corresponds 
 to a Poynting vector which does not vanish on the 
 average, and to a radiation pressure which is not exactly 
 balanced in all directions. This physical interpretation is 
 well justified in Stochastic Electrodynamics, where the 
 zero-point field is regarded as a real random field, whose 
 effects can only be observed in the presence of some cause 
 of inhomogeneity or anisotropy. As a consequence of the 
 unbalanced radiation pressure, any accelerated observer 
 feels a resistance to its acceleration. It has been shown by 
 Haisch and Rueda \cite{hr1} that this resistance is proportional 
 to acceleration and has therefore the typical property of 
 inertia. To this end, they considered a sequence of Lorentz 
 transformations between the accelerated system and local 
 co-moving inertial systems. 
  
 The idea that the inertia of matter could be the result of its 
 interaction with vacuum fluctuations is appealing from the 
 logical point of view, because in this approach it is not 
 necessary to start from an equation of motion for the 
 observer which already contains an inertial term. 
  
 In other words, while any classical Lagrangian and any 
 wave equation (or field equation in the case of second 
 quantization) is an evolution of Newton's second law $ 
 {\bf   F} = m {\bf a} $, in the approach mentioned above 
 inertia follows from the anisotropic radiation pressure. In 
 this way, inertia appears as a consequence of the third law, 
 i.e., of momentum conservation, and ultimately of 
 translation invariance. 
  
 There is, of course, a problem with the size of the 
 accelerated particle. The part of the vacuum fluctuations 
 spectrum acting upon the particle depends on its size; only 
 for finite (non-zero) size can the ``inertial effect" be finite. 
 The standard approach to this kind of vacuum effects in 
 QED is to consider point-like, 
 structureless particles, and re-absorb 
 infinite contributions to their self-mass into the ``bare mass" 
 through the renormalization procedure.  
  
 Now, it is interesting to compare the results of the 
 ``radiation pressure" approach, based only upon momentum 
 conservation, with those of standard relativistic field 
 theories with ``built in" second Newton law. What happens 
 if we introduce finite cut-offs in the field theoretical 
 expressions for the self-mass $ \Sigma $? One finds that 
 the result depends much on the spin of the particles. For 
 scalar particles, it is possible to introduce a cut-off in $ 
 \Sigma $, set the bare mass to zero and interpretate 
 somehow the physical mass as entirely due to vacuum 
 fluctuations -- except for the problem that the ``natural" cut-offs 
 admitted in QFT (supersymmetry scale, GUT scale, Planck 
 scale) all correspond to very large masses. For spin 1/2 
 particles (QED with fermions) one obtains a relation 
 between bare mass and renormalized mass which is 
 compatible both with the observed electron mass and with 
 a finite cut-off, but only if the bare mass is not zero. 
 Below we shall give the explicit expressions for the scalar 
 and spinor case. Before that, however, let us make a short
 premise and consider a semiclassical approximation.
  
 \medskip \noindent
 {\bf Effective mass of charged particles in a thermal or 
 stochastic electromagnetic field.} 
 
 Let us consider charged 
 particles with bare mass $m_0$ immersed in a thermal or 
 stochastic background $A_\mu(x)$. For scalar particles 
 described by a quantum field $\phi$, the Lagrangian density is of the 
 form 
 	\begin{equation} 
	L = \frac{1}{2} \phi^* (P^\mu - e A^\mu)
	\phi (P_\mu - e A_\mu)
	- \frac{1}{2} m_0^2 |\phi|^2
\label{lagr}
\end{equation}
 and contains a term $e^2 \phi^* A_\mu A^\mu 
 \phi$, which after averaging on $A_\mu$ can be regarded 
 as a mass term for the field $\phi$. Take, for instance, the 
 Coulomb gauge: The effective squared mass turns out to 
 be equal to $m^2 = m_0^2 + e^2 \langle | {\bf A}({\bf x}) 
 |^2 \rangle$. 
  
 For homogeneous black body radiation at a given 
 temperature $T$, the average is readily computed 
 \cite{mas}. One has 
 \begin{equation} 
 \langle | {\bf A} |^2 \rangle = 
 \int_0^\infty d\omega \frac{u_\omega}{\omega^2} 
 \label{ave} 
 \end{equation} 
 where $u_\omega$ is the Planck spectral energy density. 
 By integrating one finds that the squared mass shift is 
 given by $\Delta m^2 = const. \sqrt{\alpha} kT$ (the 
 constant is adimensional and of order 1).
 This mass shift can be significant for a hot plasma.
If the plasma is dense, the average
$\langle |{\bf A({\bf x})}|^2 \rangle$ will be itself
defined in part by fluctuations and correlations
in the charge density \cite{akh}. Therefore the mass
shift for one kind of particles in the plasma depends
in general on which other particles are present.
  
 We can also insert into eq.\ (\ref{ave}) the Lorentz-invariant 
 spectrum of vacuum fluctuations (see \cite{mil,hrklu}; 
 this is valid both for Stochastic Electrodynamics 
 and QED). In this case the integral diverges, and a 
 frequency cut-off is needed (compare our discussion 
 above). Alternatively, one can introduce in the integral
 an adimensional weight function $\eta(\omega)$ peaked at some
 ``resonance" frequency $\omega_0$. In this way, one finds that 
 $\Delta m \sim \omega_0$ in natural units
 ($\hbar=c=1$). Haisch and Rueda
 have proposed \cite{hr2} to set $\omega_0 \sim 
 \omega_{Compton}$ for the electron, so that
 $\Delta m \sim m_e$. 
  
 In conclusion, field theory offers a quite straightforward 
 method to evaluate the influence of a classical random 
 background on effective mass. This appears to work, 
 however, only for scalar particles. The Dirac Lagrangian is 
 linear with respect to the field $A_\mu$, therefore it is 
 impossible to obtain a mass term for spinors by averaging 
 over the electromagnetic field. 
 
 \medskip \noindent
 {\bf The case of QED and relation to mass renormalization.}
 
 For a quantized electromagnetic field the simple considerations
 above are not applicable. One must compute the propagator
 of the charged particles taking into account QED corrections.
 It is found that the pole of the propagator (giving the
 physical mass) is shifted, due to the effect of vacuum
 fluctuations. 
 
 Before mass renormalization, the full {\em scalar}
 propagator has the form
 	\begin{equation} 
	\Delta_0(p) = \frac{1}{p^2 - m_0^2 -
	\Sigma(m^2) + i\varepsilon} 
\end{equation} 
	where $m_0$ is the ``bare" mass and $\Sigma(p^2)$ 
represents the sum of all possible connected vacuum 
polarization diagrams. In the Feynman renormalization procedure,
which appears to be the most suitable for our purposes,
$\Sigma(p^2)$ is Taylor-expanded and the mass renormalization
condition 
	\begin{equation} 
	m_0^2 + \Sigma(m^2) = m^2 
\label{renorm}
\end{equation} 
is imposed, where $m_0$
and $\Sigma(m^2)$ are diverging quantities, but their difference
is finite.

An alternative view, corresponding to the idea of inertia
as entirely due to vacuum fluctuations, is the following:
set $m_0=0$, impose a physical cut-off $M$ in $\Sigma$
and compute $m$ from (\ref{renorm}). For scalar QED, one finds
in this way that $\Sigma$ is of the order of the cut-off.

In QED with {\em spinors} the mass renormalization condition
involves a logarithm \cite{iz}:  
	\begin{equation} 
	m-m_0 = \left[ \Sigma(p^2,M) 
	\right]_{\sqrt{p^2}=m} = m_0 \frac{3\alpha}{4\pi}
	\left( \ln \frac{M^2}{m_0^2} + \frac{1}{2}
	\right)
\label{mild}
\end{equation} 
Therefore $m_0$ must be non-zero. For an estimate, let us
rewrite the cut-off in natural units as $M=10^\xi$ $cm^{-1}$
and set $m/m_0 \equiv k >1$ (i.e., vacuum fluctuations increase
the mass by a factor $k$). Eq.\ (\ref{mild}) becomes
	\begin{equation} 
	k-1 = \frac{3\alpha}{2\pi}
	\left( \xi \ln 10 - \ln m + \ln k + \frac{1}{4}
	\right)
\end{equation}
Let us apply this to the electron ($m \sim 0.5 \cdot 10^{10}$
$cm^{-1}$), taking the Planck mass as cut-off ($\xi \sim 33$).
Solving with respect to $k$ we find that $k \sim 1.19$ -- a 
quite moderate renormalization effect, after all. For smaller
cut-offs, $k$ turns out to be closer to 1.

 \bigskip
\noindent
{\bf Acknowledgments} - This work was supported in part
by the California Institute for Physics and Astrophysics
via grant CIPA-MG7099. The author is grateful to 
V.\ Hushwater and A.\ Rueda 
for useful discussions.


\begin{thebibliography}{99}

\bibitem{sto}
L.\ de la Pena and A.\ Cetto, {\it The quantum dice:
an introduction to Stochastic Electrodynamics},
Kluwer, Dordrecht, 1996.

\bibitem{ud}
P.C.W.\ Davies, J.\ Phys.\ {\bf A 8} (1975) 609.
W.G.\ Unruh, Phys.\ Rev.\ {\bf D 14} (1976) 870.

\bibitem{hr1}
B.\ Haisch and A.\ Rueda, {\it Inertia as a reaction of
the vacuum to accelerated motion}, Phys.\ Lett.\ 
{\bf A 240} (1998) 115-126; {\it Contribution to inertial 
mass by reaction of the vacuum 
to accelerated motion}, Found.\ of Phys.\ {\bf 28} (1998)
1057-1108. 

\bibitem{mas}
G.\ Modanese, {\it Inertial mass and vacuum fluctuations in
quantum field theory}, report hep-th/0009046.

\bibitem{akh}
A.I.\ Akhiezer et al., {\it Plasma Electrodynamics},
Pergamon, New York, 1975.

\bibitem{mil}
P.W.\ Milonni, {\it The quantum vacuum}, Academic Press,
New York, 1994.

\bibitem{hrklu}
B.\ Haisch and A.\ Rueda, {\it The zero-point field
and inertia}, in {\it Causality and Locality in Modern
Physics}, 171-178, G.\ Hunter, S.\ Jeffers and J.-P.\
Vigier eds., Kluwer, Dordrecht, 1998.

\bibitem{hr2}
Y.\ Dobyns, B.\ Haisch and A.\ Rueda, {\it Inertial mass and
the quantum vacuum fields}, report gr-qc/0009036, to appear in 
Annalen
der Physik.

\bibitem{iz}
C.\ Itzykson and J.-B.\ Zuber, {\it Quantum field theory},
McGraw-Hill, New York, 1985.

\end{thebibliography}
\end{document}